\documentclass[english]{revtex4}
\usepackage[T1]{fontenc}
\usepackage[latin9]{inputenc}
\usepackage{float}
\usepackage{amsmath}
\usepackage{graphicx}
\usepackage{amssymb}
\usepackage{esint}

\makeatletter
\@ifundefined{textcolor}{}
{%
 \definecolor{BLACK}{gray}{0}
 \definecolor{WHITE}{gray}{1}
 \definecolor{RED}{rgb}{1,0,0}
 \definecolor{GREEN}{rgb}{0,1,0}
 \definecolor{BLUE}{rgb}{0,0,1}
 \definecolor{CYAN}{cmyk}{1,0,0,0}
 \definecolor{MAGENTA}{cmyk}{0,1,0,0}
 \definecolor{YELLOW}{cmyk}{0,0,1,0}
 }

\makeatother

\usepackage{babel}

\begin{document}

\title{Asymptotic predictions using short-time data in oscillating billiards}

\author{Kushal Shah}

\affiliation{Faculty of Mathematics and Computer Science, Weizmann Institute of
Science, POB 26, Rehovot 76100, Israel}

\email{shah.kushal@yahoo.com}
\begin{abstract}
Particle motion in a smoothly oscillating non-integrable billiard
is known to result in unbounded energy growth. Though the asymptotic
energy growth rate of an ensemble of particles in an oscillating chaotic
billiard is known to be quadratic, there are no estimates available
for smoothly oscillating pseudo-integrable billiards. The energy growth
rate in such systems is so slow that it is very hard to predict the
asymptotic rates from finite time computations. In this paper, a method
is proposed to estimate the asymptotic energy growth rate in a system
by using short-time data. The idea is applied to the case of an oscillating
pseudo-integrable system, and it is shown that the asymptotic energy
growth rate in such systems could be exponential.
\end{abstract}
\maketitle

\section{Introduction}

For particle motion governed by an autonomous Hamiltonian, the total
energy remains constant. If we now add a small time-dependent smooth
perturbation, the total energy is no longer independent of time. For
such systems, the energy can remain bounded or grow unboundedly with
time. For 1d autonomous Hamiltonians, one can show using the KAM theorem
\cite{Tabor,LiebLicht} that the energy will remain bounded under
the influence of a small time-dependent perturbation. Though KAM theorem
is applicable to 2d and higher dimensional integrable systems too,
the energy growth in such systems can still be unbounded due to a
phenomenon known as Arnold diffusion \cite{ArnDiff}. However, this
process is exponentially slow and hard to observe in practical situations.
For the case of chaotic Hamiltonians, it was recently shown that the
energy can grow unboundedly, at a much faster rate compared to Arnold
diffusion, if a small time-dependent perturbation is added to the
system \cite{GT_CMP}.

One could carry out a similar analysis for an important class of non-smooth
Hamiltonians, namely, the dynamical billiard. A billiard is a dynamical
system in which a particle moves in straight lines within a region
bounded by rigid boundaries, and undergoes specular reflections on
collision with the boundary \cite{Birk,Bunim,Sinai}. Particle motion
in oscillating billiards has been studied in the context of Fermi
acceleration \cite{Fermi,Ulam,Zasl}. The concept of Fermi acceleration
has found immense applications in areas like collisional heating in
plasma RF sheaths \cite{LiebGody} and models of nuclear fission \cite{Blocki}.
There also have been several studies of the Fermi accelerator from
a quantum mechanical viewpoint \cite{Jose,Makowski}. For the 1d Fermi-Ulam
model, it has been proved, using the KAM theorem, that the particle
energy remains bounded \cite{Pust}. For the 2d case, it has been
numerically shown that energy can grow unboundedly in a smoothly oscillating
billiard if the frozen system is chaotic \cite{Carvalho,Kamphorst,Koiller,Leonel,LRA}.
A theory has also been proposed to explain this observation \cite{GT_JPA}. 

In between completely integrable and completely chaotic systems, there
is an interesting class of pseudo-integrable systems. Such systems
are non-integrable but are not chaotic (all Lyapunov exponents are
zero). Liouville-Arnold theorem \cite{Arnold_Book} states that if
a Hamiltonian system with $n$ degrees of freedom has $n$ independent
integrals in involution, then it can be integrated by quadratures.
However, this notion of integrability is only local. There are systems
which have the required number of independent integrals in involution,
but do not have global action-angle variables \cite{Duist,Kozlov}.
And this is because the manifold corresponding to the particle motion
in these systems is not diffeomorphic to a sphere or a torus, but
to a surface of genus greater than one. Rational polygons belong to
this interesting class of pseudo-integrable systems \cite{Gutkin1,Gutkin2,Berry}.
A natural question arises regarding the energy growth of a particle
if a small time-dependent smooth perturbation is added to such systems.
Recently, it has been numerically shown that a particle experiences
unbounded energy growth in smoothly oscillating versions of pseudo-integrable
billiards \cite{Shah}.

For smoothly oscillating chaotic billiards, it has been numerically
observed that the rate of energy growth is quadratic in time \cite{LRA}.
And this observation is also supported by theoretical arguments \cite{GT_JPA}.
However, for the case of simply connected pseudo-integrable systems,
there are no known estimates for the rate of energy growth. This paper
is the first attempt to give some numerical estimate for the asymptotic
energy growth rate in such systems.

Carrying out numerical computations for energy growth in oscillating
billiards is computationally intensive. And the lack of accuracy after
a certain number of collisions puts a limit on the total time for
which the simulation can be carried out. This can have important implications
for estimating the asymptotic energy growth rate of particles in a
given system. Basically, if the growth rate up to a certain time has
been observed to be quadratic, there is no guarantee that a higher
growth rate will not be observed if the computations could be carried
out for a longer time. Thus, it is very important to have some indirect
way of predicting the asymptotic growth rate based on short-time data.
This paper aims at proposing one such method.

\section{Oscillating billiards}

Consider a 2d billiard with an oscillating boundary. The phase space
variables are $\left\{ x_{n},y_{n};u_{n},v_{n};t_{n}\right\} $ where
$n$ denotes the number of collisions with the oscillating boundary.
The particle energy is $E_{n}=u_{n}^{2}+v_{n}^{2}$. In general, $u_{n}$
and $v_{n}$ are not constant. For the sake of simplicity, let us
consider a billiard whose oscillating boundary is just a straight
line and moves in a direction normal to itself. Let the wall velocity
at the instant of the $n$th collision be $w_{n}$ and always along
the $y$-axis. The change of velocity of the particle in the $n$th
collision with the oscillating boundary is

\begin{equation}
\Delta v_{n}=2w_{n}\label{eq:Delta_v}\end{equation}
The time interval between two collisions with the oscillating boundary
is \begin{equation}
\Delta t_{n}=\frac{l_{n}}{\sqrt{E_{n}}}\label{eq:Delta_t}\end{equation}
where $l_{n}$ is the distance travelled by the particle in between
the $n$th and $(n-1)$th collision. As particle energy grows unboundedly
($E_{n}\rightarrow\infty$), the time-interval between collisions
decays to zero ($\Delta t_{n}\rightarrow0$), and in this limit, we
have\begin{equation}
\frac{dv}{dt}=\sqrt{E}\frac{2w}{l}\label{eq:dv/dt}\end{equation}
Substituting $v=\sqrt{E}\cos\phi$, we get\begin{equation}
\frac{dE}{dt}=E\left[2\tan\phi\frac{d\phi}{dt}+\frac{4w}{l\cos\phi}\right]\label{eq:dE/dt}\end{equation}
Now, what we are interested in is the growth rate averaged over the
time period of oscillations of the base. Time averaging Eq. \eqref{eq:dE/dt},
we get\begin{equation}
\frac{d\overline{E}}{dt}=\overline{E}\left\langle 2\tan\phi\frac{d\phi}{dt}+\frac{4w}{l\cos\phi}\right\rangle _{T}\label{eq:E_average_eqn}\end{equation}
where $T=2\pi\big/\omega$ is the time period of oscillations of the
billiard. For the 1d Fermi-Ulam model, $\cos\phi=1$, and $l$ is
a known periodic function if $E$ is high enough. Thus, in this case,
the average of the right hand side of Eq. \eqref{eq:E_average_eqn}
is zero and there is no net acceleration of the particle. However,
for a general 2d billiard, $l$ and $\cos\phi$ change randomly, and
if the average, $\left\langle 2\tan\phi\frac{d\phi}{dt}+\frac{4w}{l\cos\phi}\right\rangle _{T}>0$,
then this leads to an unbounded growth of energy. If the long time
average,\begin{equation}
\lim_{t\rightarrow\infty}\frac{1}{t}\int_{0}^{t}\left[2\tan\phi\frac{d\phi}{dt}+\frac{4w}{l\cos\phi}\right]\mbox{d}t=R>0\label{eq:long_time_average}\end{equation}
then the particle also experiences exponentially fast growth of energy,
$\overline{E}=E_{0}e^{Rt}$. However, for a given system, if $R$
is very low, then it is hard to see the asymptotic exponential growth
numerically. This is mainly because numerically accurate computations
can be done only for a finite time, which may be too short to observe
the exponential energy growth in some systems. In this paper, a way
of estimating asymptotic growth rates using short-time data is proposed.

\section{Making asymptotic predictions}

The most general equation for evolution of energy of an ensemble of
particles in any system is\begin{equation}
\frac{dE}{dt}=f\left(E,t\right)\label{eq:dE/dt_General}\end{equation}
where $f$ is an arbitrary function of $E,t$. The method proposed
in this paper is applicable to a sub-class of Eq. \eqref{eq:dE/dt_General},
namely, those equations where $f\left(E,t\right)$ is variable separable,
\begin{equation}
\frac{dE}{dt}=f\left(E,t\right)=g\left(E\right)h\left(t\right)\label{eq:dE/dt_gh}\end{equation}
where $g\left(\cdot\right)$ and $h\left(\cdot\right)$ are two arbitrary
functions of their arguments. As can be seen from Eq. \eqref{eq:E_average_eqn},
the oscillating billiard certainly belongs to this class of systems.
If both the functions $g,h$ are known analytically, then one can
surely go ahead and do rigorous analysis. However, in certain situations
like particle motion in oscillating billiards, the function, $h(t)$,
has terms that change in a way that is random for all practical purposes.
In such situations, one cannot do any rigorous analysis for predicting
the system behavior. Also, as mentioned earlier, one also cannot do
computations for very long times due to limits on the accuracy. 

The basic idea proposed in this paper is that measuring the dependence
of short-time growth rates on the initial energy of the system can
give us some indication of asymptotic behavior. For various asymptotic
energy growth rates, it can be shown, using Eq. \eqref{eq:dE/dt_gh},
that the dependence of the linear part of the growth rate on initial
energy, $E_{0}$, must be\begin{eqnarray}
\mbox{Linear growth rate} & : & E=E_{0}+\alpha t\nonumber \\
\mbox{Quadratic growth rate} & : & E=E_{0}+\beta E_{0}^{0.5}t+...\nonumber \\
\mbox{Exponential growth rate} & : & E=E_{0}+\gamma E_{0}t+...\label{eq:LinQuadExp_GrowthRate}\end{eqnarray}
where $\alpha,\beta,\gamma$ are arbitrary constants. Thus, the idea
is to carry out the simulations till a time for which the growth rate
is predominantly linear and then, using Eq. \eqref{eq:LinQuadExp_GrowthRate},
we can say that
\begin{enumerate}
\item If a plot of $\left(E-E_{0}\right)$ is independent of $E_{0}$, the
asymptotic growth rate is linear
\item If a plot of $\left(E-E_{0}\right)\big/E_{0}^{0.5}$ is independent
of $E_{0}$, the asymptotic growth rate is quadratic
\item If a plot of $\left(E-E_{0}\right)\big/E_{0}$ is independent of $E_{0}$,
the asymptotic growth rate is exponential
\end{enumerate}
Now, we use this idea to estimate the asymptotic energy growth of
an ensemble of particles in an oscillating trapezium (see Fig. \ref{enu:Geometry-of-trapezium}),
which is a pseudo-integrable system. For the case of a trapezium with
an oscillating base, Fig. \ref{enu:Polygon_Exp} shows that $\left(E-E_{0}\right)\big/E_{0}$
is independent of $E_{0}$ (approximately), hence indicating that
the asymptotic energy growth rate in this billiard could be exponential.
In Fig. \ref{enu:Polygon_Quad}, a plot of $\left(E-E_{0}\right)\big/E_{0}^{0.5}$
can be clearly seen to depend on $E_{0}$. Though the plots in Figs.
\ref{enu:Polygon_Exp} and \ref{enu:Polygon_Quad} are not conclusive,
they do provide a reasonable indication of the asymptotic energy growth
rate being exponential. 

\begin{center}
\begin{figure}[H]
\begin{centering}
\includegraphics[angle=270,scale=0.5]{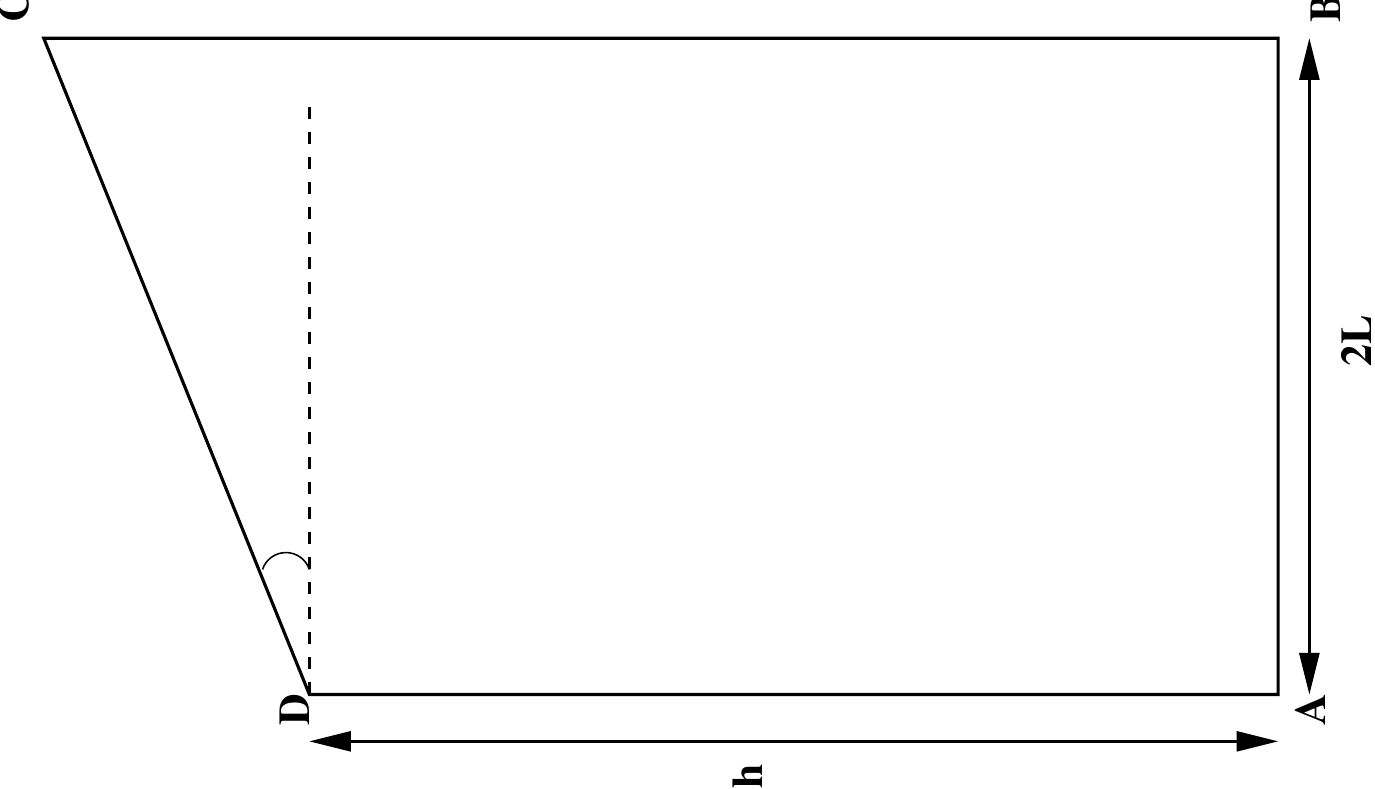}
\par\end{centering}

\caption{Geometry of the trapezium. }
\label{enu:Geometry-of-trapezium}
\end{figure}

\par\end{center}

\begin{center}
\begin{figure}[H]
\begin{centering}
\includegraphics[scale=0.4]{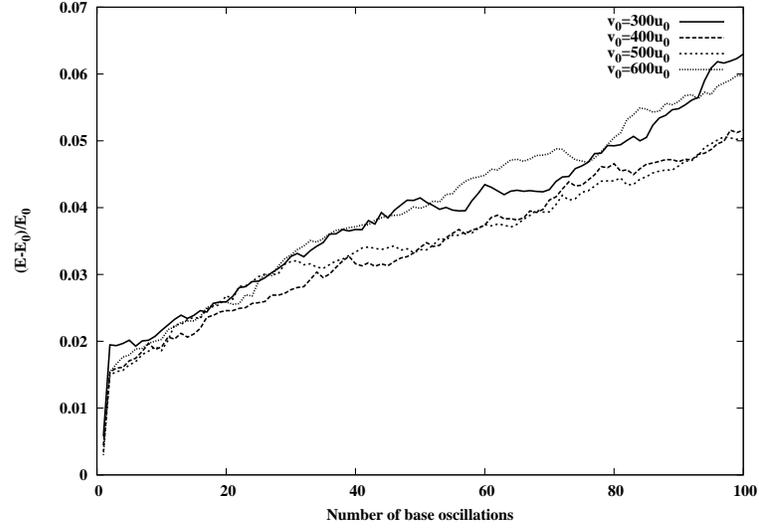}
\par\end{centering}

\caption{Plot of $\left(E-E_{0}\right)\big/E_{0}$ vs. time for an ensemble
of 1000 particles in an oscillating trapezium with parameters: $L=0.5$,
$h=4.0$ and $\angle ADC=0.5\pi+\pi\big/18$. Amplitude of oscillations
of the base (AB) was $0.1$ with a frequency of $\omega=0.1$, and
with an initial horizontal velocity, $u_{0}=4\omega\big/\pi$ . }
\label{enu:Polygon_Exp}
\end{figure}

\par\end{center}

\begin{center}
\begin{figure}[H]
\begin{centering}
\includegraphics[scale=0.4]{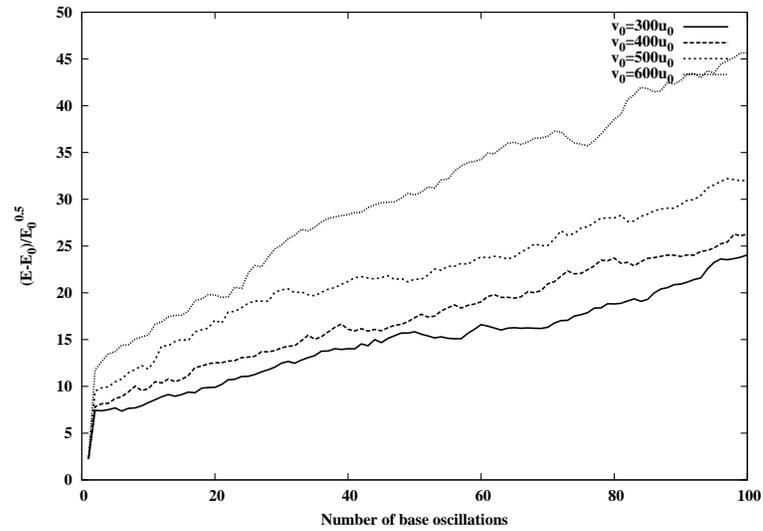}
\par\end{centering}

\caption{Plot of $\left(E-E_{0}\right)\big/E_{0}^{0.5}$ vs. time for an ensemble
of 1000 particles in an oscillating trapezium with the same parameters
as in Fig. \ref{enu:Polygon_Exp}. }
\label{enu:Polygon_Quad}
\end{figure}

\par\end{center}

\section{Conclusions}

Though the idea presented in this paper has been applied to the specific
case of an oscillating billiard, it is very general in nature. There
are many problems in engineering and applied sciences where prediction
of asymptotic behavior is of crucial importance. One limitation of
the method proposed in this paper is that, using this method, it is
hard to differentiate between two growth rates that are not well separated,
eg. exponential versus a polynomial of high order. However, I hope
that the idea proposed in this paper will provide a starting point
in developing more elaborate tools to handle this problem.
\begin{acknowledgments}
I acknowledge support of the Israel Science Foundation and the Minerva
Foundation. I would also like to thank Vered Rom-Kedar and Dmitry
Turaev for interesting discussions.\end{acknowledgments}

\end{document}